\documentclass{stacs_proc}   
\usepackage{amsthm, amsmath, amssymb}
\usepackage{pifont, color}

\newcommand{\sha}{\, {\mathbin{\mathchoice
{\rule{.3pt}{1ex}\rule{.3em}{.3pt}\rule{.3pt}{1ex}
\rule{.3em}{.3pt}\rule{.3pt}{1ex}}
{\rule{.3pt}{1ex}\rule{.3em}{.3pt}\rule{.3pt}{1ex}
\rule{.3em}{.3pt}\rule{.3pt}{1ex}}
{\rule{.2pt}{.7ex}\rule{.2em}{.2pt}\rule{.2pt}{.7ex}
\rule{.2em}{.2pt}\rule{.2pt}{.7ex}}
{\rule{.3pt}{1ex}\rule{.3em}{.3pt}\rule{.3pt}{1ex}
\rule{.3em}{.3pt}\rule{.3pt}{1ex}}\mkern2mu}} \,}

\newcommand{\ra}{\rightarrow}

\newcommand{\N}{\mathbb{N}}
\newcommand{\Q}{\mathbb{Q}}
\def\L{{\mathcal{L}}}
\def\D{{\mathcal{D}}}
\def\E{{\mathcal{E}}}

\def\PP{{\mathcal P}}

\begin{document}
\title[Analytic aspects of the shuffle product]{Analytic aspects of
  the shuffle product}
\author[MJM]{M. Mishna}{Marni Mishna}
\address[MJM]{Department of Mathematics, Simon Fraser University, Burnaby, Canada}
\email{mmishna@sfu.ca}

\author[MZ]{M. Zabrocki}{Mike Zabrocki}
 \address[MZ]{Department of Mathematics and Statistics, York University, Toronto, Canada}
\email{zabrocki@mathstat.yorku.ca}

\begin{abstract}
  There exist very lucid explanations of the combinatorial origins of
  rational and algebraic functions, in particular with respect to
  regular and context free languages. In the search to understand how
  to extend these natural correspondences, we find that the shuffle
  product models many key aspects of D-finite generating functions, a
  class which contains algebraic. We consider several different takes
  on the shuffle product, shuffle closure, and shuffle grammars, and
  give explicit generating function consequences. In the process, we
  define a grammar class that models D-finite generating functions.
\end{abstract}
\maketitle

\stacsheading{2008}{561-572}{Bordeaux}
\firstpageno{561}

\section*{Introduction}
\subsection*{Generating functions of languages}
The (ordinary) generating function of a language~$\L$ is the sum
\[
L(z)=\sum_{w\in\L} z^{|w|},
\]
where $|w|$ is the length of the word. This sum is a formal power
series if there are finitely many words of a given length. In this
case, we say the language is {\em proper}, and we can rewrite~$L(z)$
as~$L(z)=\sum \ell(n) z^n$, where $\ell(n)$ is the number of words in
$L$ of length $n$. In the case where we have an unambiguous grammar to
describe a regular language or a context free language, one can
automatically generate equations satisfied by generating function
directly from the grammar. These are the well known translations:
\[\begin{array} {rlcrl}
\L &= \L_1+\L_2 & \implies & L(z)&=L_1(z)+L_2(z)\\
\L&=\L_1\cdot\L_2& \implies & L(z)&=L_1(z)L_2(z)\\
\L&=\L_1^* & \implies &L(z)&= (1-L_1(z))^{-1}.
\end{array} \]
Generating functions of formal languages are now a very established
tool for algorithm analysis (see~\cite{FlSe06} for many references)
and increasingly for random generation~\cite{DuFlLoSc04}.  In this
context, we are also
interested in the {\em exponential generating function\/} of a
language. The two are related by the Laplace-Borel transform, however
it is sufficient for our purposes to think of the exponential
generating function $\hat{L}(z)$ as the Hadamard product of $L(z)$ and
$\exp(z)=\sum \frac{z^n}{n!}$; that is,
$\hat{L}(z)=\sum\ell(n)\frac{z^n}{n!}$.

  One spectacular feature of generating functions of languages is the
  extent to which their analytic complexity models the complexity of the
  language.  Specifically, we have the two classic results: first,
  regular languages have rational generating functions, and second,
  those context-free languages which are not
  inherently ambiguous have an algebraic
  generating function. The context-free languages form a large and
  historically important subclass of all objects which have algebraic
  generating functions.  Bousquet-M\'elou provides us ~\cite{MBM05,
    MBM06} with an interesting discussion of the nature of
  combinatorial structures that possess algebraic and rational
  generating functions, including broad classes that are not
  representable as context-free languages.

  There remain unanswered questions related to other classes of
  languages, and other classes of functions. An example of the former
  is the question of Flajolet~\cite{Flajolet87}: ``In which class of
  transcendental functions do generating functions of (general)
  context free languages lie?''  An example of the latter is the
  identification of languages whose generating functions are {\em
    D-finite}\footnote{D-finite, also known as holonomic, functions
    satisfy linear differential equations with polynomial
    coefficients.}. This is an exceptional class of
  functions~\cite{Stanley99}, which, for the moment, lacks a
  satisfying combinatorial explanation. We survey some current
  understandings in Section~\ref{sec:dfin}, and provide a language
  theoretic interpretation of one in Section~\ref{sec:term-point}.

  To capture the analytic complexity of D-finite generating functions we
  should not expect a simple climbing of the language hierarchy (to
  indexed or context sensitive, say), as there are different notions of
  complexity in competition. For example the language $\{a^nb^nc^n:
  n\in \N\}$ is difficult to recognize, but trivial to
  enumerate. Likewise, the generating function of the relatively
  simple looking language $\{z^{n^2}: n\in \N\}$ has a natural
  boundary at $|z|=1$, which is a trademark of very complex analytic
  behaviour.

\subsection*{The shuffle product}
In the absence of the obvious answers, we consider
a very common, and useful operator, the {\em shuffle product}, and
discover that it fills in many interesting holes in this
story. Consider the words $w$,~$uw_1$ and~$vw_2$, and the letters $u, v\in\Sigma$.
We define the shuffle product of  two words recursively by the equation
\[
u w_1\sha v w_2 = u(w_1\sha vw_2) + v(uw_1 \sha w_2), 
\quad w\sha\epsilon=w;\quad \epsilon \sha w =w.
\]
 Here the union is disjoint, and we distinguish duplicated
 letters from the second word by a bar: $a\sha a = \{\overline{a}a, a\overline{a}\}$.
Using the shuffle product we can define a class of languages with
associated generating functions that form a class that strictly
contains algebraic functions; it allows us to model a very
straightforward combinatorial interpretation of the derivative (indeed
in some interesting non-commutative algebras the shuffle product is
even called a derivative); and it allows us to neatly consider some
larger classes which are simultaneously more complex from the
language and generating function points of view.

\subsection*{Goal and Results}
The aim of this study is two-fold.  We hope that a greater understanding of
generating function implications of adding the shuffle product 
to context free languages provides insight to a larger class of
combinatorial problems.
The second goal is to understand the combinatorial interpretations of
different function classes that arise between algebraic and
D-finite. The shuffle is a natural combinatorial product to consider
since it is, in some sense, a generalization of pointing.

In the present work, we  first examine the shuffle as an operator {\em on\/}
languages, and in the second part we consider the shuffle as a grammar
production rule {\em to define\/} languages. We show that
the shuffle closure of the context free languages is D-finite; we give
the asymptotic growth of coefficients of two classes using shuffle; we define
a special pointing class that describes all D-finite functions; and
discuss the shuffle closure of a language.  

In the next section we review interpretations of differential
equations. This is followed by a discussion on the shuffle of
languages, and some descriptions of shuffle grammars.

\section{Interpreting differential equations combinatorially}
\subsection{The class of D-finite functions}
The class of D-finite functions is of interest to the combinatorialist for many
reasons. The coefficient sequence of a D-finite power series is
P-recursive: it satisfies a linear recurrence of fixed length
with polynomial coefficients, and hence is easy to generate,
manipulate, and even ``guess''  their form. By
definition, D-finite functions satisfy linear differential equations with polynomial
coefficients, and thus it is relatively straightforward in many cases to
perform an asymptotic analysis on the coefficients, even without a
closed form for the generating function. One important feature that we
use here is that a P-recursive sequence grows asymptotically like 
\[\ell(n)\sim\lambda(n!)^{r/s}\exp(Q(n^{1/m}\omega^nn^\alpha(\log n)^k))\]
where $r, s, m,n ,k\in\N$, $Q$ is a polynomial and  $\lambda, \omega, \alpha$,
are complex numbers.  We contrast this to the asymptotic template
satisfied by coefficients of algebraic functions:
\begin{equation}\label{eqn:alg-asympt}
\ell(n) \sim \kappa\frac{n^{d}}{\Gamma(d+1)}\omega^{-n},
\end{equation}
where $\kappa$ is an algebraic number and $d\in \Q\setminus \{-1, -2,
\dots\}$. (A very complete source on the theory of asymptotic
expansions of coefficients of algebraic functions arising in the
combinatorial context is~\cite[Section VII.4.1]{FlSe06}.)
Notable differences include the exponential/ logarithmic factors, the
power of a factorial, and the allowable exponents of $n$.

We shall use the following properties of the D-finite
functions: The function $1/f$ is D-finite if, and only if, $f$ is of the
  form $\exp(g) h$, where $g$ and $h$ are algebraic~\cite{Singer86}; 
The Hadamard product $f\times g=\sum f_n g_n z^n$ of two D-finite functions $f=\sum f_n z^n$ and $g=\sum g_n z^n$ is also D-finite.

\subsection{The simplest shuffle: the point}
Pointing (or marking) is an operation that has been long studied in
connection with structures generated by grammars. The point of an word
$w$, denoted $P(w)$, is a set of words, each with a different position
marked. For example, $P(abc)=\{ \overline {a}bc, a\overline{b}c,
ab\overline{c} \}$. From the enumerative point of view we remark that
the two languages $L$, and $\L_1=P(\L)=\{P(w): w\in \L\}$, satisfies
the enumerative relation 
\begin{equation}\label{eqn:diff}
\ell_1(n)= n\ell(n),
\end{equation} and hence
$L_1(z)=z\frac{d}{dz}L(z)$. The pointing operator is relevant to our discussion
because of the simple bijective correspondence between $P(\L)$ and
$\L\sha a=\{w\sha a: w\in \L \}$.

The first obvious question is, ``does pointing increase expressive
power?''.  In the case of regular languages and context free languages
the answer is no; We can add a companion non-terminal for each non
terminal that generates a language isomorphic to the pointed language. 
Let $\overline{A}$ be the pointed version of $A$. We add the following rules which model pointing:
\[\overline{(AB)}= \overline{A}B+A\overline{B}, \quad \overline{(A+B)}=\overline{A}+\overline{B}\]
Remark how these rules resemble the corresponding product and sum
rules for differentiation.  Furthermore, from the point of view of
generating functions, we know that the derivative of a rational
function is rational again, and the derivative of an algebraic
function is again algebraic, and so we know immediately that we could
not hope to increase the class of generating functions represented by
this method.

Pointing, when paired with a ``de-pointing'' operator which
removes such marks, becomes powerful enough to describe other kinds of
constructions, namely labelled cycles and sets~\cite{FlZiVa94,
  Greene83}. In this case we can describe set partitions, and
which has exponential generating function $\exp(\exp(z)-1)$, which is
not D-finite. 

It takes much more effort~\cite{BoFuKaVi07} to define a pointing
operator with a differentiation property as in Eq.\eqref{eqn:diff}
for unlabelled structures defined using Set and Cycle
constructions. It is a fruitful exercise, as one can then generate
approximate size samplers with expected linear time complexity.
 
\subsection{Other combinatorial derivatives}
\label{sec:dfin}
Combinatorial species theory~\cite{ BeLaLe98} provides a
rich formalism for explaining the interplay between analytic and
combinatorial representions of objects. In particular, using the
vehicle of the the cycle index series, and there are several
possibilities on how to relate them to (multivariate) D-finite
functions~\cite{LaLa07,Mishna07}. In this realm, given any arbitrary
linear differential equation with polynomial coefficients we can
define a set of grammar operators that allow us to construct a pair of
species whose difference has a generating function that satisfies the
given differential equation. Unfortunately at present we lack the
intuition to understand what this class ``is'', specifically, we lack
the tools to construct a test to see if any given class or language
falls within it.

In Section~\ref{sec:shuf-closure} we give a language theoretic interpretation of the
derivative of a species; specifically a grammar system, from which,
for any linear differential equation with coefficients from $\Q[x]$ we
can generate a language whose generating function satisfies this
equation.
\subsection{Other differential classes}
There are several other natural function classes related to the
differential equations. A series $f(z)\in K[[t]]$ is said to be {\em constructible
  differentiably algebraic (CDF)\/} if it belongs to some finitely
generated ring which is closed under differentiation.~\cite{BeRe90,
  BeSa95}. This is equivalent to satisfying a system of differential
equations of a given form.  Combinatorially, any CDF function can be
interpreted as a family of enriched trees. Theorem 3 of ~\cite{BeRe90}
gives the result that if $\sum a_n/n! t^n$ is CDF,
then $|a_n|=O(\alpha^n n!)$ for some complex constant $\alpha$. This
class is not closed under Hadamard product, and any arbitrary CDF
function is unlikely to have the image under the Borel transform also
CDF.  This is the key closure property required for a meaningful
correspondence with respect to the shuffle product.
 
A larger class which contains both CDF and D-finite is differentiably
algebraic.  A function is {\em differentiably algebraic (DA) } if it satisfies an 
algebraic differential equation of the form $P(x, y,  y', \dots,
y^{(n)})=0$ where $P$ is a non-trivial polynomial in its $n+2$
variables. (See Rubel's survey~\cite{Rubel89} for many references.) 

The set of DA functions is closed under multiplicative inverse and
Hadamard product. These two facts together are sufficient to prove
that {\em all of the classes we consider are differentiably algebraic}.

\subsection{Generating functions and shuffles}
Generating functions are useful tool for the automatic studies of
certain combinatorial problems. The shuffle operator has a
straightforward implication on the generating function, as we shall
see.

With the aid of the shuffle product, Flajolet {\em et al.}
\cite{FlGaTh92} are able to perform a straightforward analysis of four
problems in random allocation.  By using some systematic translations,
they are able to derive integral representations for expectations and
probability distributions.
As they remark, the shuffle of languages appears in several places
relating to analysis of algorithms (such as evolution of two stacks in a
common memory area). 

\section{The shuffle of two languages}
The shuffle of two languages is defined as 
\[
\L_1 \sha \L_2 =\bigcup_{w_1\in \L_1,w_2\in \L_2} (w_1 \sha w_2). 
\] 
In order to use a generating function approach, we assume that $\L_1$
is a language over the alphabet $\Sigma_1$, and $\L_2$ is a language
over $\Sigma_2$, and $\Sigma_1\cap\Sigma_2=\emptyset$. If they share
an alphabet, it suffices to add a bar on top of the copy from $\Sigma_2$. 

\subsection{The shuffle closure of context free languages}
\label{sec:shuf-closure}
We consider the shuffle closure of a language in the next section, and
first concentrate on the shuffle closure of a class of languages. For
any given class of languages ${\mathcal C}$, the shuffle closure can
be defined recursively as the (infinite) union of $S_0, S_1, \dots$,
the sequence recursively defined by
\[ S_0 =\mathcal{C}, \qquad S_n= \{\L_1 \sha \L_2:\L_1\in S_{n-1},\L_2
\in \mathcal{C}\}.\] The shuffle product is commutative and
associative~\cite{Lothaire83}, and thus the closure contains $S_j\sha
S_i$, for any $i$ and $j$. Remark, that for any given language in the
closure, there is a bound on the number of shuffle productions that
can occur in any derivation tree; namely, if $\L\in S_n$, that bound
is $n$. 

In general, we denote the closure of a class of languages under
shuffle as ${\mathcal{C}}^{\sha}.$ The class of regular languages is
closed under the shuffle product, since the shuffle of any two regular
languages is regular. However, the context free languages are not
closed under the shuffle product~\cite{Lothaire83}, and hence we
consider its closure.  

The prototypical language in this class is the shuffle of (any finite number
of) Dyck languages. Let $|w|_a$ be count the number of occurences of
the letter ~$a$ in the word~$w$. Let $\D$ be the Dyck language over the alphabet
$\Sigma=\{u, d\}$:
\[
\D = \{ w\in \Sigma^*: w'v=w \implies |w'|_u \geq |w'|_d \text{ and } |w|_u=|w|_d.\}
 \]
We construct an isomorphic version $\E$, over the alphabet
$\{l, r\}$. 

The language $\D\sha \E$ has encodes random walks restricted to the
quarter plane with steps from u(p), d(own), r(ight), and l(eft) that
return to the origin. By considering the larger language of Dyck
prefixes, we can models walks that end anywhere in the quarter plane. Indeed, as
the shuffle does preserve two distinct sets of prefix conditions, there
are many examples of random walks in bounded regions that can be
expressed as shuffles of algebraic languages.

It might be interesting to consider other standard questions of
classes of languages for this closure class; in particular if
interesting random walks arise. 
\subsection{The closure is D-finite}
In order to show that the shuffle product of two languages with
D-finite generating functions also has a D-finite generating function,
we consider the following classic observation on the enumeration of
shuffles of languages. 

If $\L$ is the shuffle of $\L_1$ and $\L_2$, then the number of  words of length
$n$ in $\L$ are easily counted if the generating series for
$L_1(z)=\ell_{1}(n)z^n$ and $L_2(z)=\ell_{2}(n)z^n$ are known by the
following formula:
\[
\ell(n) = \sum_{n_1+n_2=n}\binom{n}{n_1\, n_2}\ell_1(n_1)\ell_2(n_2).
\] 
To see this, recognize that a word in $\L$ is a composed of two
words, and a set of positions for the letters in the word from $\L_1$,
This is equivalent to 
\begin{equation}
\frac{\ell(n)}{n!} = \sum_{n_1+n_2=n}\frac{\ell_1(n_1)}{n_1!}\frac{\ell_2(n_2)}{n_2!},
\end{equation}
which amounts to the relation between the {\em exponential generating
  functions} of the three languages:
\begin{equation}\label{eqn:shuf_df}
\L=\L_1\sha\L_2  \qquad\implies\qquad \hat{L}(z)=\hat{L}_1(z)\hat{L}_2(z).
\end{equation}
Using these relations, we can easily prove the following result. 
\begin{proposition}
If $\L_1$ and $\L_2$ are languages with D-finite ordinary generating functions,
then the generating series for $\L=\L_1\sha \L_2$, $L(z)$ is also D-finite.
\end{proposition}
As is the case with many of the most interesting closure properties of
D-finite functions, the proof follows from the closure of D-finite
functions under Hadamard product~\cite{Lipshitz88}.  \proof Since
D-finite functions are closed under Hadamard product, the ordinary
generating function is D-finite if and only if the exponential
generating function of a sequence is  D-finite. Consequently, if
$L_1(z)$ and $L_2(z)$ are D-finite, then so are the exponential
generating functions, $\hat{L_1}(z)$ and $\hat{L_2}(z)$. By closure
under product, $\hat{L}(z)$ is D-finite, and thus so is $L(z)$.  \qed
This result has the following consequences.
\begin{corollary}
If $\L_1$ and $\L_2$ are context free languages which are not inherently
ambiguous, then the generating series $L(z)$  for $\L=\L_1\sha \L_2$ is
D-finite. 
\end{corollary}

\begin{corollary}
Any language in the shuffle closure of context free languages has a
D-finite generating function. 
\end{corollary}

\subsection{Asymptotic template for $\ell(n)$}
We continue the example from the previous section using the two Dyck
languages $\D$ and $\E$. It is straightforward to compute that
$D(z)=E(z)=\sum \binom{2n}{n}\frac{1}{n+1} z^n$. Thus, $\ell(n)$, the
number of words of length $n$ in the shuffle is given by
\[
\ell(n)=\sum_{n_1+n_2=n}\binom{n}{n_1}\binom{n_1}{n_1/2}\binom{n_2}{n_2/2}.
\]
We remark that an  asymptotic expression for $\ell(n)$ can be
determined by first using the Vandermonde-Chu identity to simplify $\ell(n)$:
\[
\ell(n)= \binom{n}{\lfloor n/2\rfloor}\binom{n+1}{\lceil n/2 \rceil},
\] 
and then by applying Stirling's formula. Since $\ell(n)\sim 4^n/n$, we
see that it the resulting series is not algebraic. Flajolet uses this
technique extensively in~\cite{Flajolet87} to prove that certain
context-free languages are inherently ambiguous.  Thus, we have that
our class has generating functions strictly contains the algebraic
functions.

Thus, we have some elements of a class of function with a nice
asymptotic expansion. A rough calculation gives that that the shuffle of two
languages, with respective asymptotic growth of $\kappa_i
n^{r_i}(\alpha_i)^n$, for $i=1,2$ respectively, is given by the
expression
\[\ell(n)\sim\kappa n^{r_1+r_2}(\alpha_1+\alpha_2-r_1-r_2)^n.\]
 How could one hope to prove directly that all elements in this class
have an expansion of the form \[\ell(n)\sim \kappa \alpha^n n^r,\]
where now $r$ can be {\em any\/} rational, and $\kappa$ is no longer
restricted to algebraic numbers? It seems that it should be possible
to prove this at least for the shuffles of series which satisfy the
hypotheses of Theorem 3.11 ~\cite{MBM06}, using a more generalized
form of the Chu-Vandermonde identity, or for the closure of the
sub-class of context-free languages posessing an $\N$-algebraic
generating function. In this case the $d=-3/2$, and this simplifies
the analyses considerably.  Unfortunately, it does not seem like a
direct application of Bender's method~\cite[Theorem VI.2]{FlSe06}
applies.

Theorem~\ref{thm:nfactorial} states that the
asymptotic form will not contain any powers of $n!$ greater than
2. This illustrates a limitation with the expressive power of the
shuffle closure of context free languages: there
are known natural combinatorial objects which have D-finite generating functions
with coefficients  that grow asymptotically with higher powers of
$n!$. For example, the number of  $k$-regular graphs for $k>4$
contains $(n!)^{5/2}$, and the conjectured asymptotic for for
$k$-uniform Young tableaux~\cite{ChMiSa05} contains $n!^{k/2-1}$.

\section{Shuffle grammars}
We extend the first approach by allowing the shuffle to come into play
earlier in the story; we add the shuffle operator to our grammar
rewriting rules. Shuffle grammars as defined by 
Gischer~\cite{Gischer81} include a shuffle rule, and a shuffle
closure rule.  We consider these in Section~\ref{sec:shuf-closure}.

As we did earlier, we first consider languages which have a natural
bound on the number of shuffle productions that can occur in a
derivation tree of any word in the language. That is followed by an
example of a recursive shuffle grammar to illustrate how powerful they
can be. It has been proven~\cite{Jedrzejowicz88} that the recursive
shuffle grammars do indeed have a greater expressive power, but it is
not always clear how to interpret the resulting combinatorial
families. We begin with a second kind of pointing operator. 
\subsection{A terminal pointing operator}
\label{sec:term-point}
The traditional pointing operator can be used to model
$z\frac{d}{dz}$, but one can show that this is, in fact, insufficient
to generate all D-finite functions. To remedy this, we define a
pointing operator which  mimics the concept behind the derivative of a
species. This pointing operator has the effect of converting a letter
to an epsilon by `marking' the letter. Consequently, a letter can not be marked more than
once, and each subsequent time a word is marked, there is a counter on
the mark which is augmented. The pointing operator applied
a set of words will be the pointing operator applied to each of the elements of the set.
Notationally, we distinguish them with
accumulated primes.  We give some examples:
\begin{align*}
\PP( aab ) &= a' a b + a a' b + a a b'\\
\PP( \PP( aab)) &= a' a'' b + a' a b'' + a'' a' b + a a' b'' + a'' a b' + a a'' b'\\
\PP( a''' a' b'') &= \emptyset.
\end{align*}
The length of the word is the number of unmarked letters in a word (but the combinatorial
objects in the language encode more than just the length in some sense). The
number of words in the pointing of a word is equal to its length.  

This gives a straightforward interpretation of the derivative:
\[\L_1=\PP(\L) \qquad \implies \qquad L_1(z) = \frac{d}{dz}L(z).\]
Using this definition if $A$ is a symbol which `yields' through a grammar a language

Remark, if we allow concatenation after marking,
we could generate two letters in the same word marked with a single prime
via concatenation of marked words.

Using the marking operation, we can express most D-finite functions,
specifically, by the differential equations that they satisfy.  For
example, the series $P(z)=\sum_{n \geq 0} n! z^n$ satisfies the
differential equation
\[P(z) = 1 + z P(z) + z^2 P'(z).\]
This is modelled by the grammar
\begin{align*}
A &\rightarrow \varepsilon\\
A &\rightarrow a A\\
A &\rightarrow b c \PP(A).
\end{align*}
An alphabet on three letters ($a,b,c$) allows us to track the origin
of each letter. Here is the result of the third iteration of the rules: 
\begin{align*}
1\oplus a &\oplus aa+ba'c \oplus aaa+abca'+bca'a+bcb''ca'+bcaa'+bcbc''a'
 \oplus aaaa+aabca'+abca'a.
\end{align*}

We will call a pointing grammar 
one that has rules of the form
\begin{equation}\label{genrules}
A \rightarrow w,\qquad
A \rightarrow w B,\qquad
A \rightarrow \PP(B).
\end{equation}
Despite the fact that we allow only {\em left}
concatenation, (a strategy to avoid concatenating pointed words) these
grammars rules can model any D-finite function.

We can define a procedure for finding a language given a defining
equation satisfied by a
D-finite generating function.  Say that a generating
function $T(z)$ satisfies
\begin{equation}\label{mydeq}
T(z) = q(z) + q_0(z) T(z) + q_1(z) T'(z) + \ldots + q_n(z) T^{(n)}(z)~.
\end{equation}

Now substitute $T(z) = P(z) - N(z)$ and 
$$(P(z) - N(z)) = q(z) + q_0(z) (P(z) - N(z)) + q_1(z) (P'(z) - N'(z)) + \ldots 
+ q_n(z) (P^{(n)}(z) - N^{(n)}(z))$$

Use also the notation that $q_i(z) = q_i^+(z) - q_i^-(z)$ where $q_i^+(z)$ are the positive terms
of the polynomial and $q_i^-(z)$
are the negative ones.

Then if
\begin{equation}\label{posterms}
P(z) = q^+(z) + q_0^+(z) P(z) + q_0^-(z) N(z) + \cdots + q_n^+(z) P^{(n)}(z) + q_n^-(z) N^{(n)}(z)
\end{equation}
and
\begin{equation}\label{negterms}
N(z) = q^-(z) + q_0^-(z) P(z) + q_0^+(z) N(z) + \cdots + q_n^-(z) P^{(n)}(z) + q_n^+(z) N^{(n)}(z)
\end{equation}
then
$P(z) - N(z)$ satisfies equation \eqref{mydeq}.

Now we can define a language with a rule for each monomial in \eqref{posterms} and
\eqref{negterms}
and every terms $x^a R^{(k)}(z)$ is represented by a rule of the form
$${\tilde R} \rightarrow w \PP( \cdots \PP(R) \cdots)$$
where $\PP$ occurs $k$ times and $R$, $\tilde R$ are symbols 
representing a language whose generating function is either
$P(z)$ or $N(z)$ and $w$ is a word of length $a$.

Any language which is generated from rules of the form Eq.~\eqref{genrules}
has a generating function which satisfies a linear differential
equation, and hence is D-finite. 

We summarize this in the following theorem.
\begin{theorem}
A language which is generated from the rules of the form Eq.~\eqref{genrules}
has a D-finite generating function.  Moreover, any D-finite function
can be written as a difference of two generating functions for
languages which are generated by rules of this form.
\end{theorem}
\subsection{Acyclic shuffle dependencies}
We consider languages generated by the following re-writing rules,
where $w$ is a word, and $A$, $B$ and $C$ are non-terminals:
\begin{equation}\label{eq:x}\\
A \rightarrow w, \qquad
A \rightarrow BC,\qquad
A \rightarrow B\sha C.
\end{equation}
For any language generated by rules of the above type, and a fixed set
 of non-terminals, we construct the graph with non-terminals as
nodes, and for every production rule $A\rightarrow B\sha C$, we make
an edge from $A$ to $B$ and an edge from $A$ to $C$. If this graph is
acyclic, we say the language has acyclic shuffle dependencies. The
next section treats languages that have a cyclic dependency. 

We prove that this class of languages is larger than those generated
by the pointing operator of the previous section, because we can
generate a language with a generating function that is not D-finite.  

We re-use the Dyck languages $\D$ and $\E$ defined in
Section~\ref{sec:shuf-closure}. Consider the language generated by the following grammar:
\begin{align*}
A&\ra \D\sha \E\\
C&\ra 1| AC.
\end{align*}
The shuffle dependency graph is a tree, and thus this is in our
class.  The generating functions of~$A$ and~$C$ are given by
\[
A(z)=
\frac{-1}{4z}+\frac{(16z-1)}{2\pi z}\operatorname{EllipticK}(4\sqrt{z})
+\frac{1}{\pi z}\operatorname{EllipticE}(4\sqrt{z}),\quad
C(z)=\frac{1}{1-A(z)}.
\]
Since $1-A(z)$ is not of the form $\exp(algebraic) algebraic$, $C(z)$
is not D-finite. Nonetheless, we can prove an asymptotic result about
generating functions in this class. 

\begin{theorem}\label{thm:nfactorial}
  Let $L$ be a proper language generated by shuffle production in an
  unambiguous grammar of with rules of the form given in
  Eq.~\eqref{eq:x}, on an alphabet with $k$ letters. The number of
  words of length $n$, $\ell(n)$, satisfies $\ell(n)=O(n!^2)$.
\end{theorem}
\proof
  Since the grammar generates proper languages, there are no shuffle
  productions with epsilon.  Thus, the derivation tree of a word of
  length $n$ can have at most $n$ shuffle productions. In the worst
  case, each one increments the alphabet and so the maximum size of alphabet
  that a word of length $n$ can draw on is then $kn$. The total number
  of words from this alphabet is $(kn)^n$.
   
  For $k<n$ the result follows by Stirling's formula.
\qed
\subsection{Cyclic shuffle dependencies}
Languages in this class will have an infinite alphabet since we use a
disjoint union in our shuffle. However, the number of words
of a given length is finite if there is no derivation tree possible
that is a shuffle and an $\epsilon$. Under this restriction,  any word
of length $n$  comes from an alphabet using no more than more than
a constant multiple of $n$ letters. We consider an important class of
this type in the next section.

\subsection{The shuffle closure of  a languages}
\label{sec:shuf-closure}
A class of languages which falls under this category are those that
are generating using the shuffle closure operator. The {\em shuffle
  closure\/} of a language is defined recursively in the following
way: $\L^{\sha 1}=\L\sha\L$, and $\L^{\sha n}=\L^{\sha n-1}\sha
\L$. The shuffle closure, is the union over all finite shuffles:
\[\L^{\sha}=\bigcup_n \L^{\sha n}.\]  Equivalently, we write this as a grammar
production: $ A \rightarrow A\sha B | B. $ The shuffle closure ~\cite{Janzen85,
  Jedrzejowicz88} provides extremely concise notation. In particular,
they arise in descriptions of sequential execution histories of
concurrent processes.

Remark, that the closure of the language is one single language, whereas
the closure of the class of languages that is one language is an
infinite set of languages.

The shuffle closure of a single letter gives all permutations:
\[a^{\sha}=a \oplus \overline{a}a +a \overline{a} \oplus
\overline{\overline{a}}\overline{a} a
+\overline{\overline{a}}a \overline{a} 
+ a \overline{\overline{a}}\overline{a}
+\overline{a}\overline{\overline{a}}a
+\overline{a}a\overline{\overline{a}}
+a\overline{a}\overline{\overline{a}}\oplus \dots
\]

The generating function of the this language is $\sum n!z^n$, and
indeed the generating function of the shuffle closure of any word of
length $k$ is $\sum (kn)! (\frac{z^k}{k})^{n}$, which is also
D-finite. 


To prove our formula above, we express the generating function of
$\L^{\sha}$ in terms of the operators which switch between the ordinary
and exponential generating functions. Recall, $L(z)=\sum a_n z^n\implies \hat{L}(z)=\sum
\frac{a_n}{n!},$ and we define the Laplace operator $\mathcal{L}\cdot  \hat{L}(z) =L(z)$.
Then, 
\begin{equation}\label{eqn:shufclosure}
\L_1=\L^{\sha} \qquad \implies \qquad L_1(z)=\sum_n\mathcal{L}\cdot [
(\hat{L}(z))^n]. 
\end{equation}
Although all of the summands are D-finite, it is possible that the sum
is not. 

Clearly, the shuffle closure does not preserve regularity, and indeed
adding it, and the shuffle product to regular languages is enough to
generate all recursively enumerable languages. Thus, we see that if
there is no bound on the number of shuffles possible in any expression
tree, the languages can get far more complex.

Nonetheless the following conjecture seems reasonable, and perhaps it
is possible to prove it following starting from
Eq.~\eqref{eqn:shufclosure}, and necessarily a more sophisticated
analysis.
\begin{conj}
The shuffle closure of a regular language has a D-finite generating function.
\end{conj}

\vskip-0.3cm
\section{Conclusion}
A next step is to adapt the Bolzmann generators to these
languages. Since we can effectively simulate labelled objects in an
unlabelled context, we can easily describe objects like strong
interval trees. This approach might allow a detailed analysis of
certain parameters of permutation sorting by reversals, as applied to
comparative genomics~\cite{BeBeChPa07}.

We are also interested in characterizing the context-free languages
whose shuffle is not algebraic, and to consider the other naturual
questions of closure that are standard for language classes.

\small
\subsubsection*{Acknowledgments}
We gratefully acknowledge many discussions from the Algebraic Combinatorics Seminar
at the Fields Institute. In particular, we
acknowledge contributions by
N. Bergeron, C. Hollweg, and M. Rosas. We
wish to also acknowledge the financial support of NSERC. 

\def\polhk#1{\setbox0=\hbox{#1}{\ooalign{\hidewidth
  \lower1.5ex\hbox{`}\hidewidth\crcr\unhbox0}}}

\end{document}